\documentclass[12pt,twoside]{article} 

\usepackage{graphicx}

\date{December 19, 2012} 

\begin{document} 


\centerline{} 

\centerline{} 

\centerline {\Large{\bf Excited Nuclei in neutron star crusts}} 

\centerline{} 



\centerline{\bf {N. Takibayev}} 

\centerline{Institute of Experimental and Theoretical Physics,} 

\centerline{Al-Farabi Kazakh National University, Almaty 050071, Kazakhstan} 

\centerline{} 




\centerline{\bf {Kiyoshi Kato}} 

\centerline{Nuclear Reaction Data Centre, Faculty of Science, Hokkaido University,
Sapporo 060-0810, Japan} 

\centerline{} 





\centerline{\bf {D. Nasirova}} 


\centerline{Abai Kazakh National Pedagogical University, Almaty 050010, Kazakhstan} 

\centerline{}

\begin{abstract} 
The paper considers the chains of successive electron capture reactions by nuclei of the iron group which take place in the crystal structures of neutron star envelopes. It is shown that as a result of such reactions the daughter nuclei in excited states accumulate within certain layers of neutron star crusts. The phonon model of interactions is proposed between the excited nuclei in the crystalline structure, as well as formation of highly excited nuclear states which emit neutrons and higher energy photons.
\end{abstract} 



\section{Introduction} 
Neutron stars and processes in them are of considerable interest because such stars, as well as white dwarfs and black holes, are classified as compact stellar objects and are considered to be the final stage of star life. Neutron stars are balanced by equality between the gravity forces, the compression factor, and the pressure of a degenerate gas in the stars interior, the expansion factor. Neutron stars have masses of $1\div 3$ of solar mass and radii about $10 \: km$. They can have the atmosphere about few centimeters; a kilometer-thick crystalline crusts, the mantle and central cores. Also, there are magnetospheres that can accelerate particles \cite{TK_ST,TK_HPY}. 

In this paper, the reactions and processes that occur with nuclei in crystalline structure of neutron star crusts are considered. The majority of these reactions are well studied in laboratory experiments, and others can be assessed applying fundamental laws of quantum theory, the theory of nuclear and photonuclear reactions, the degenerate electron Fermi liquid, the state of matter under extreme conditions, etc. 

We begin our consideration from outer crust layers where reactions of electron capture started by creation of neutron rich nuclei. Mass of nuclei, their angular momentum and spin, as well as excitation levels can be taken the same as in the laboratory experiments \cite{TK_Atlas,TK_NNDC}.

As usually, we took the substance in the crusts, where the density $\rho\geq 10^{5}g\cdot cm^{-3}$ , as an overdense crystalline structure  which is most energetically favorable at such extreme pressures \cite{TK_KDA}. This Coulomb crystal consisting of bare nuclei and the electron Fermi liquid will provide electrical neutrality of matter in general, and will experience huge gravitational pressure \cite{TK_ST,TK_HPY}. 

The outer crust, where $10^{6}g\cdot cm^{-3} \leq \rho \leq 10^{11}g\cdot cm^{-3}$, is the area where electron capture reactions begin. The inner crust, where $10^{11}g\cdot cm^{-3} \leq \rho \leq 10^{14}g\cdot cm^{-3}$ , is determined by neutron dripping from neutron-rich nuclei. Then in the mantle, where $ \rho \geq 2\cdot 10^{14}g\cdot cm^{-3}$   is already above the normal nuclear density, the nuclei are deformed creating unfamiliar forms \cite{TK_HPY}.

Our analysis will covers middle and lower layers of the outer crust and upper layers of the inner crust within $10^{6}g\cdot cm^{-3} \leq \rho \leq 10^{12}g\cdot cm^{-3}$ . We assume in the uppermost layers of the outer crust the element composition of the material is close to the original, which was at the center of a supernova during its explosion, when the neutron star was formed. There are basically the iron group elements. But the element composition of the material should change in depth within the neutron star envelopes because of consecutive reactions of electron capture by nuclei and related processes. 

Below in Section 2 we consider chain reactions of electron capture by nuclei. It is shown that in the middle and lower layers of the outer crust the daughter nuclei can not only exist in the ground states, but also in the excited states.

In Section 3 we propose a model of the excitation transfer between the nuclei in the overdense crystal and the formation of highly excited nuclei capable of high energy gamma  radiation and emission of free nucleons. The model is based on the representation of the phonon excitations in nuclei and in the crystal itself. 

In Section 4, we give a general description of the phonon excitations in the nuclei and their nonlinear interactions, and interactions between  nuclear excitations and excitations of the crystal lattice. 

The conclusion also contents the assumption that the gas of free neutrons already exists in the middle and lower layers of the outer crust, and not just in the inner crust. It can significantly enrich and complicate the dynamics of the interactions in the envelopes of neutron stars

\section{Electron capture reactions} 

The density of matter in the layers of neutron star outer crust can be associated with the Fermi energy of the electrons $E_{F}=\sqrt{p^{2}_{F}c^{2}+m^{2}_{e}c^{4}}$  by the expression: $\rho = (A/2Z)\cdot 1.948\cdot 10^{6} x^{3} g\cdot cm^{-3}$, where $x = p_{F}/m_{e}c$ ,  $p_{F}$ is the Fermi impulse, $\lambda_{e} = \hbar/m_{e}c$  is the Compton wavelength of electron, $A$ is the mass number, $Z$ is the charge of nuclei \cite{TK_ST}. 
 
Capture of electrons by free protons start at  $\rho\geq 1.2\cdot 10^{7}g\cdot cm^{-3}$, but each nuclide has its own threshold energy and threshold density, i.e. the threshold layer where the reaction of electron capture can start.  

Each stable nuclide will therefore give own consistent chain of electron capture reactions. The daughter nuclei arising in these reactions, which decay under normal terrestrial conditions, will persist as stable nuclei in the overdense matter. They cannot emit an electron due to the electron Fermi liquid countering. Moreover, in many reactions the daughter nuclei will emerge both in the ground and excited states. Below we show the chains of reactions generated by the stable isotopes of iron group \cite{TK_MSU, TK_HU}. 
 
For iron isotope $^{54}Fe$, whose abundance in nature is $5,845 \% $, one can write the following chain of reactions:
\begin{eqnarray}
^{54}Fe + e^{-} \rightarrow ^{54}Mn + \nu \ , \ \ \ \ \ \ E_{e} > 0.697 \ MeV \ , \nonumber \\
^{54}Mn + e^{-} \rightarrow ^{54}Cr + \nu \ , \ \ \ \ E_{e} > -1.377 \ MeV \ , \\   			
^{54}Cr + e^{-} \rightarrow ^{54}V + \nu \ ,  \ \ \ \ \ \ E_{e} > 7.042 \ MeV \ , \nonumber \\
... \ \ \ \ \ \ \ \ \ \ \ \ \ \ \ \ \ \ \ \ \ \ \ \ \ \ \ \ \ \ \ \ \ \ \ \ \ \ \ \ \ \ \ \ \ \ \ \ \ \ \ \ \ \ \ \  \nonumber  	
\end{eqnarray}
The threshold energies for capture reactions are specified here on the right. The threshold density for $^{54}Fe$   would be $\rho_{th} \approx 1.555\cdot 10^{6}g\cdot cm^{-3}$ . Note that the electron capture by this nuclide will occur at lower densities than even the electron capture by a proton.

It is noticeable that the threshold energy for the second reaction is lower than the energy of the first reaction for about $2 \ MeV$. That means, the second reaction for $^{54}Mn$   is open and exoenergetic.  In the corresponding threshold layer, chromium nuclei will be generated in both ground and excited states: $E^{*}_{1}(^{54}Cr) = 0.835 \ MeV$   and   $E^{*}_{2}(^{54}Cr) = 1.824 \ MeV$. Thus, in the second reaction in the chain (1), there are three open channels:
\begin{eqnarray}
^{54}Mn(3^{+}) + e^{-} \rightarrow ^{54}Cr(0^{+}) + \nu \ , \nonumber \\
^{54}Mn(3^{+}) + e^{-} \rightarrow ^{54}Cr^{*}_{1}(2^{+}) + \nu \ , \\   			
^{54}Mn(3^{+}) + e^{-} \rightarrow ^{54}Cr^{*}_{2}(4^{+}) + \nu \ .  \nonumber 	
\end{eqnarray}
Here in brackets after the each nucleus the quantum number specifies the level that the nucleus occupies.

It is known that the transitions with the smallest change of the quantum numbers have an advantage in the reactions of weak decay, for example, with the smallest change of the angular momentum between the initial and the final states of nuclei. Moreover, increase of the difference by one unit only leads to the suppression of the probability of the corresponding transitions for three or more orders of magnitude. We accept the same rule should be applicable for the reverse reactions also. It then follows that nuclei in the excited states are mainly formed in reactions (2).    

Note that the third reaction in the chain (1) can start in the deeper layers of the outer crust,  where $\rho \geq \rho_{th3} \approx 5.06\cdot 10^{9}g\cdot cm^{-3}$ , and the nucleus $^{54}Cr$  can capture electrons. 

Let us now consider a stable isotope of iron $^{56}Fe$ , whose abundance in nature is $91,754 \% $. For this nucleus the chain of reactions is as follows:
\begin{eqnarray}
^{56}Fe + e^{-} \rightarrow ^{56}Mn + \nu \ , \ \ \ \ \ \ E_{e} > 3.695 \ MeV \ , \nonumber \\
^{56}Mn + e^{-} \rightarrow ^{56}Cr + \nu \ , \ \ \ \ \ \ E_{e} > 1.629 \ MeV \ ,\nonumber \\	
^{56}Cr + e^{-} \rightarrow \ \ ^{56}V + \nu \ ,  \ \ \ \ \ \ E_{e} > 9.201 \ MeV \ ,  \\
^{56}V + e^{-} \rightarrow \ ^{56}Ti + \nu \ , \ \ \ \ \ \ \ E_{e} > \ 7.14 \ MeV \ ,\nonumber \\
^{56}Ti + e^{-} \rightarrow ^{56}Sc + \nu \ , \ \ \ \ \ \ \ E_{e} > 13.64 \ MeV \ , \nonumber \\
^{56}Sc + e^{-} \rightarrow ^{56}Ca + \nu \ , \ \ \ \ \ \ \ E_{e} > \ 11.9 \ MeV \ , \nonumber \\
... \ \ \ \ \ \ \ \ \ \ \ \ \ \ \ \ \ \ \ \ \ \ \ \ \ \ \ \ \ \ \ \ \ \ \ \ \ \ \ \ \ \ \ \ \ \ \ \ \ \ \ \ \ \ \ \  \nonumber  	
\end{eqnarray}

	So, at densities $\rho \geq 7.155\cdot 10^{9}g\cdot cm^{-3}$  the first reaction starts and the daughter nuclei $^{56}Mn$ are generated, which under normal terrestrial conditions are unstable. For these nuclei the second reaction should be already open since its threshold is nearly for $2 \ MeV$ below the threshold of the first reaction. The energies of the two lowest excited states of nucleus equal: $E^{*}_{1}(^{56}Cr) = 1.007 \ MeV$   and   $E^{*}_{2}(^{56}Cr) = 1.832 \ MeV$. 
	The reactions with nuclei excitation are therefore also possible:
\begin{eqnarray}
^{56}Mn(3^{+}) + e^{-} \rightarrow ^{56}Cr(0^{+}) + \nu \ , \nonumber \\
^{56}Mn(3^{+}) + e^{-} \rightarrow ^{56}Cr^{*}_{1}(2^{+}) + \nu \ , \\   			
^{56}Mn(3^{+}) + e^{-} \rightarrow ^{56}Cr^{*}_{2}(4^{+}) + \nu \ .  \nonumber 	
\end{eqnarray}
Here, as in reactions (2), the majority of generated nuclei will be in excited states.  
 
The third reaction in the chain (3) will open at even greater depths where $\rho \geq \rho_{th3} \approx 1.132\cdot 10^{10}g\cdot cm^{-3}$. Remarkable that in the second cascade of reactions the fourth reaction is opened immediately with the third because its threshold is less than the threshold of the previous reaction for more than $2 \ MeV$ again.

Fifth reaction of chain (3) can be realized at even greater depths when $\rho \geq \rho_{th5} \approx 3.697\cdot 10^{10}g\cdot cm^{-3}$. And like in the first two pairs, the sixth reaction in the chain (3) will be immediately opened with the fifth reaction. Here, the threshold energy of sixth reaction is less than the threshold energy of the fifth reaction for $1.74 \ MeV$.

In contrast to the even nuclei, the isotope $^{57}Fe$  gives a different picture. $^{57}Fe$  abundance in nature is $0.754 \%$. The chain of capture reactions is as follows:
\begin{eqnarray}
^{57}Fe + e^{-} \rightarrow ^{57}Mn + \nu \ , \ \ \ \ \ \ E_{e} > 2.693 \ MeV \ , \nonumber \\
^{57}Mn + e^{-} \rightarrow ^{57}Cr + \nu \ , \ \ \ \ \ \ E_{e} > 4.963 \ MeV \ ,\nonumber \\	
^{57}Cr + e^{-} \rightarrow \ \ ^{57}V + \nu \ ,  \ \ \ \ \ \ E_{e} > 8.334 \ MeV \ ,  \\
^{57}V + e^{-} \rightarrow \ ^{57}Ti + \nu \ , \ \ \ \ \ \ \ E_{e} > \ 10.69 \ MeV \ ,\nonumber \\
... \ \ \ \ \ \ \ \ \ \ \ \ \ \ \ \ \ \ \ \ \ \ \ \ \ \ \ \ \ \ \ \ \ \ \ \ \ \ \ \ \ \ \ \ \ \ \ \ \ \ \ \ \ \ \ \  \nonumber  	
\end{eqnarray}
Here every subsequent reaction has the energy threshold greater than that of the previous one, i.e. the reactions in this chain shall take place one by one deeper to the crust.	

Let us give a chain of reactions to another stable isotope of iron $^{58}Fe$ . The abundance of this isotope in nature is $0.282 \%$.
\begin{eqnarray}
^{58}Fe + e^{-} \rightarrow ^{58}Mn + \nu \ , \ \ \ \ \ \ E_{e} > 6.323 \ MeV \ ,  \\
^{58}Mn + e^{-} \rightarrow ^{58}Cr + \nu \ , \ \ \ \ \ \  E_{e} \ > \ 4. \ \ \ MeV \ . \ \nonumber 	  	
\end{eqnarray}

Here again, the subsidiary reaction is already open, as its threshold is lower than that of the first reaction. This forms possibilities for additional reaction channels to form nuclei in excited states at the output.

Note that a similar situation exists for the isotopes of $Ni$ and other elements of the iron group. In particular, for $^{58}Ni$ with its abundance in nature $68 \%$, the chain of reactions  should be:
\begin{eqnarray}
^{58}Ni + e^{-} \rightarrow ^{58}Co + \nu \ , \ \ \ \ \ \ E_{e} \ > \ \ 0.381 \ MeV \ ,  \\
^{58}Co + e^{-} \rightarrow ^{58}Fe + \nu \ , \ \ \ \ \ \  E_{e}  > \ - 2.307 \ MeV \ , \nonumber 	  	
\end{eqnarray}
continued by the reactions as in (6).

\section{Exited Nuclei in Overdense Crystalline Structures} 

Excited nuclei arising in the crystalline structures in the envelopes of neutron stars should manifest themselves in unusual ways compared to their behavior under ordinary terrestrial density.

Thus, at terrestrial conditions the transition from the lowest excited state of the nucleus $^{54}Cr^{*}_{1}$  to its ground state emits a gamma with an energy of $E_{\gamma} \approx 835 \ keV$ and the wavelength  $\lambda_{\gamma}\approx 1.5\cdot 10^{3} \ fm$.

The distances between the nuclei in conventional crystals are $d > A^{\circ} $ , i.e. for two or three orders are larger. Excited nuclei have therefore no difficulties emitting gamma rays. Further proliferation of gamma rays would be stipulated by scattering processes in the media.

Quite a different situation occurs in the overdense crystals. Indeed, for the threshold density of the first reaction in (1), the parameter of overdense lattice is equal to $ d \approx 6.7\cdot 10^{3} \ fm$ , i.e. it is a little larger than the wavelength of radiation from an excited nucleus $^{54}Cr^{*}_{1}$. But in the chains of reactions with $^{56}Fe$ and $^{58}Fe$ the lattice parameters are significantly smaller than the wavelength of gamma rays from the excited states. Therefore, emission from excited nuclei within the lattice is either difficult or even impossible.

So, in an overdense lattice we have a complicated situation when under compression the threshold of electron capture reaction is attained and formation of  excited states of nuclei becomes possible, but this lattice does not allow the same excited nuclei emit gamma in the usual way. However, there should exist the ways out of excited states for the nuclei in the super compressed lattice, even unconventional ones. And in this paper we discuss the possibilities for their occurrence. 

We propose a model based on interrelated processes that lead to the release of nuclear excitation energy. The model is associated with peculiarities of the compressed crystalline medium in the first place, with the ordered position of nuclei (it is almost a perfect Coulomb crystal), with the presence of a degenerate electron Fermi liquid, phonon excitations in the lattice and in the nuclei, and with the nature of the internal excitations in nuclei.

The model can be represented by the following block diagram: 
\begin{figure}[h]
	\includegraphics[width=14cm]{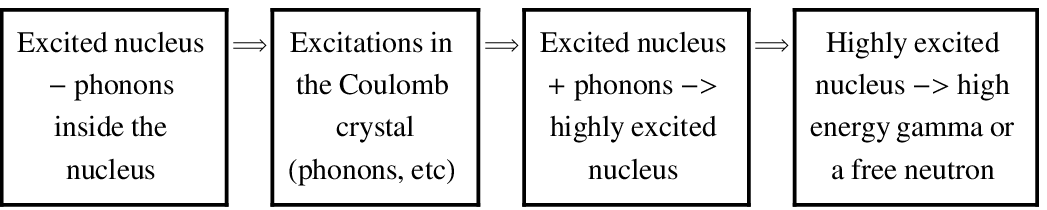}
	\caption{Flow diagram for the consequent processes in an overdense crystal.}
	\label{fig:1}
\end{figure}

The model assumes that the excitations from the nucleus transfer (transform) into excitations in the crystalline lattice, and then transfer to another excited nucleus. Additional excitation leads to the fact that the excited nucleus becomes highly excited. Then, the highly excited nucleus can emit a gamma of higher energy or give off a nucleon.
 
In the first block of Fig.1 the excited nucleus is considered as a structure with phonons inside. A nucleus, as a system of nucleons bound by the effective field, acquires excitations under external impacts or reactions; these excitations can be described by nuclear phonons \cite{TK_SVG}. The interaction of phonons can produce virtual photons, and these pass again into phonons.  
 
Second block denotes excitation already in the Coulomb crystal in the border area with the excited nucleus. Such excitations can be phonons and excitons or polaritons, Fermi electrons, 
etc. \cite{TK_BK,TK_KL}. The model assumes that the surface oscillations of the nucleus can lead to excitations and fluctuations in the Coulomb lattice. For example, the process can be represented as a virtual photon transition in the crystal lattice phonons and excitons \cite{TK_SVG,TK_BK}.

It is important that the excitations can spread in the crystal. Their interaction with the neighboring nucleus in the Coulomb lattice can cause the reverse process, for example, the transition of crystal phonons and excitations to the virtual photon and nuclear phonons inside the neighboring nucleus. If the neighboring nucleus was already in an excited state, greater excitation of the nucleus can happen, i.e. formation of the nucleus in a highly excited state \cite{TK_SVG}. This process is illustrated in the third block at Fig. 1.

The block 4 shows the final result  -  possible reactions for the highly excited nucleus. These can be emissions of a high-energy photon, a free nucleon from the nucleus such as emission of a neutron, or knocking out of nucleon from the nucleus by high-energy gamma quantum. 

Note that the formation of strong excitation in the nucleus is a nonlinear process \cite{TK_SVG}. This process can be regarded similarly to the phenomenon of high harmonic generation in nonlinear optics where a medium with high polarizability and high pumping rate of electromagnetic waves are required for the effect to take place. In our model, the nucleus itself is a medium with nonlinear response and high polarizability, and the pumping rate of elementary nuclear excitations is assured by the high number density of excited nuclei. 

High harmonic generation is a well-known phenomenon in nonlinear optics. In laboratory experiments with strong laser pumping the modes of high multiplicity (more than $2000$)have already been achieved \cite{TK_SSS, TK_Gan}. And this has been achieved on Earth.  Surely, nonlinear effects should be even more powerful in the overdense matter of neutron star envelopes. 

\section{Phonons in the nuclei and in the crystal lattice} 

Let us assume following \cite{TK_SVG} that the excited states of nuclei are caused by collective motion of nucleons in the nucleus, what results in a periodic time dependence of the nuclear properties. Vibrational excitations of the nuclei appear as a series of discrete levels at excitation energies below the threshold energy of nucleon emission $E^{*} < 7 \ MeV$.  At higher energies, the vibrational excitation of the nuclei are observed in shape of wide resonance peaks in the cross sections of nuclear reactions (giant resonances) \cite{TK_SVG}.

Let the density of nuclear matter in the ground state is $\rho(r)$ at the point $r$, then the vibrational excitations are described by $\delta \rho (r,t)$ ; any fluctuation can be represented a combination of the normal vibrational modes. For the normal modes of a spherical nucleus $\delta \rho (r,t) = \delta \rho_{L}(r)\cdot Y_{L,M}(\theta, \varphi)\cdot cos(\omega_{L} t)$ , where $\delta\rho_{L}$  describes density variation as a function of radius $r$;$Y_{L,M}(\theta, \varphi)$ is a spherical function, the indices $L$ and $M$ correspond to different types of vibrations.  

	In the quantum description each mode corresponds to the vibrational quanta - phonons.  Vibrational excitations of nuclei are characterized by the number of phonons $n_{LM}$; each phonon has the angular momentum $L$ and energy $\hbar \omega_{L}$, parity $\pi = (-1)^{L}$. Due to the incompressibility of nuclear matter the density changes at shape oscillations are primarily associated with the surface of the nucleus. 

The phonon model gives the equidistant spectrum of $n$-phonon states with energies $E_{nL} = n \hbar \omega_{L}$ for each mode. Electromagnetic transitions between the levels are to obey certain selection rules and intensity relations. Thus, for states with parallel "aligned" angular momenta of $n$ phonons the probabilities of $n \rightarrow n-1$ transitions increase in the $n$-fold compared to the $1\rightarrow 0$  transition  from the one-phonon state to the ground state (like in laser amplification [8,10]). 

The triplet levels of $4^+, 2^+, 0^+ $  with energy of $E \approx 2 \hbar \omega $  is typical for the quadrupole modes. Namely such angular momenta are possible at addition of the quadrupole moments of two phonons, and these are the characteristic levels for the considered excited nuclei in reactions (2) and (4).

Let us now consider the phonons in the crystal lattice itself, i.e. in the outer region with respect to the excited nucleus. We believe that the wave of excitation of nuclear matter inside the nucleus should, at the interface, generate phonons  in the crystal lattice and the electronic excitations in the Fermi liquid, such as excitons. Recall that in the ordinary crystals, the excitation of the atoms (or molecules) is transferred by excitons sequentially from one atom to another, so that the quantum of excitation is transferred to macroscopic distances \cite{TK_BK, TK_KL}.  

The same way nuclear excitation can be passed by excitons from one nucleus to another in the overdense crystals. Interaction with the lattice phonons allows "non-radiative" transitions at removal of excitation in the nuclei \cite{TK_BK}. 

One can consider oscillations in the nucleus based on different models. For example, a wave of nucleus excitation can be considered as a virtual photon, and a model of photon-phonon interactions, including multiphonon processes, can be applicable. If the wave vector and the energy of photon and phonon are the same, then due to this interaction, the conversion of photon into a phonon becomes possible. 

Below we estimate the wave vector of the reciprocal lattice. For $\rho \geq 7.155\cdot 10^{9}g\cdot cm^{-3}$ the distance between nuclei in the lattice is $d \leq 288 \ fm$ , and the corresponding wave number $K \geq 2.18\cdot 10^{11} \ cm^{-1}$ . Energy of the emitted gamma with such wavelength would be equal to $4.3 \ MeV$, i.e. energies of the lowest levels of the excited nuclei are to be within the first Brillouin zone and resonance overlapping of nuclear and phonon energy levels becomes possible. 

The quantum mechanical consideration involves creation and annihilation operators \cite{TK_Kit}. For example, the Hamiltonian can be written as 
\begin{eqnarray}
H = \sum_{\vec{k},\lambda} \left\{ck\left(a^{+}_{\vec{k},\lambda}a_{\vec{k},\lambda} + \frac{1}{2} \right) + \omega_{L}(1 + 4\beta \pi)^{\frac{1}{2}} \left(b^{+}_{\vec{k},\lambda}b_{\vec{k},\lambda} + \frac{1}{2} \right)   \right\} +  \nonumber \\
+ \ i\sum_{\vec{k},\lambda}\left\{\left[\frac{\pi c k \beta \omega_{L}}{(1 + 4\beta \pi )^{\frac{1}{2}}} \left(a^{+}_{\vec{k},\lambda} b_{\vec{k},\lambda} - a_{\vec{k},\lambda} b^{+}_{\vec{k},\lambda} - a_{-\vec{k},\lambda} b_{\vec{k},\lambda} + a^{+}_{-\vec{k},\lambda} b^{+}_{\vec{k},\lambda}\right)\right] \right\} \ ,
\end{eqnarray}
where $\beta = \chi/\omega^{2}_{L} $, $\chi$ is the susceptibility of the medium. 

The first term in the second sum in (8) corresponds to the photons' birth at disappearance of the phonon and the second term -  vice versa. The two following terms describe the simultaneous production or destruction of both the photon and phonon. 

The Hamiltonian is diagonalized by introducing annihilation operators $\alpha_{\vec{k}}=\delta\rho_{L} a_{k} + x b_{k} + y a^{+}_{-k} + z b^{+}_{-k}$
and the coefficients are chosen so as to satisfy the ratio $\left[\alpha_{\vec{k}}, H \right] =\omega_{k}\alpha_{\vec{k}}$.

It is important to note that the excited nuclei generated in the electron capture reactions, cannot undergo into the ground state by emitting gamma since the medium completely reflects the corresponding electromagnetic waves. The number of excited nuclei in the overdense lattice will therefore increase over time, and the nonlinear effects and interactions will expand.

Nonlinear interactions take place not only in a nucleus, but also between the excited nuclei in the lattice. Their nonlinear interaction can be fostered by both the crystal lattice itself and the overlapping of the wave functions of neighboring excited nuclei. Here the wave synchronism and coherency of excited nuclear states can be provided by the tunnel effect between such states. 

\section{Conclusion} 

So, we can summarize that the excited nuclei in the outer crust generated in the reactions of electron capture with nuclides of iron with even mass numbers such as  $^{54}Fe, ^{56}Fe, ^{58}Fe$ or $^{58}Ni, ^{62}Ni$, cannot simply emit gamma quanta within the overdense lattice.

We believe that the answer is in the process of nonlinear interaction of excited nuclear states and formation of highly excited states in these nuclei. It is obvious that nonlinear interactions will increase sharply with increasing of the number density of excited nuclei. 

When the density of excited nuclei reaches a critical value, the nuclei in highly excited states will appear. Their energies can be much higher than the excitation energy at a conventional electron capture. One can say that this is an analogue to high harmonic generation for the case of polarized media subjected to intense laser pumping.

This means that the energy of a highly excited nucleus is already enough to trigger a series of reactions ranging from the emission of high harmonics gamma rays and subsequent production of neutrino-antineutrino pairs, to the free nucleon emission from the highly excited nucleus. At that, protons should transform to neutrons due to electron capture reactions.

Thus, in the envelopes of the neutron star, gas of free neutrons can occur even in the lower layers of outer crust at $\rho \geq 7.155\cdot 10^{9}g\cdot cm^{-3}$; or even in the layers above if we take into account the chains of reactions with $^{54}Fe$ and $^{58}Ni$. Presence of free neutrons in the outer crust can cause a number of other reactions and effects \cite{TK_Tak1, TK_Tak2, TK_Tak3}, consideration of which may be important to describe the state of matter in the envelopes of neutron stars. Note, for example, that free neutrons can resonantly interact with other nuclei. The resonance range extends from the neutron resonance energies of a few $keV$ to several hundred $keV$ to form a forest of resonant states \cite{TK_Atlas}. These reactions can also generate nuclei in the excited states. \\

This project IPS 1133/GF is supported by the Ministry of Education and Sciences of the Republic of Kazakhstan. \\

{\bf ACKNOWLEDGEMENTS.} The authors are grateful the participants of the International  Conference "Nuclear Science and Its Application", Samar- kand, Uzbekistan, for useful discussions, also thank M.Abishev, A. D'Adda, N. Kawamoto and H. Quevedo for valuable recommendations.


\begin{thebibliography}{99}
\bibitem{TK_ST}
{S.~Shapiro, S.~Teukolsky, \em Black Holes, White Dwarfs, and Neutron Stars,} United States: John Wiley and Sons, 1983. 
\bibitem{TK_HPY}
{P.~Haensel, A.Y.~Potekhin, D.G.~Yakovlev, \em Neutron Stars,} Kluwer Academic Publishers, 2007. 
\bibitem{TK_Atlas}
{S.F. Mughabghab, \em Atlas of Neutron Resonances,} Elselvier BV, Amsterdam, 2006.
\bibitem{TK_NNDC}
{Nuclear Wallet Cards, USA National Nuclear Data Center - NNDC,} URL:http://www.nndc.bnl.gov/wallet/wccurrent.html. 
\bibitem{TK_MSU}
{The database "Nucleus Ground State Parameters", Moscow State University,} http://cdfe.sinp.msu.ru/
\bibitem{TK_HU}
{Nuclear Reaction Data Centre, Hokkaido University,} http://www.jcprg.org/
\bibitem{TK_KDA}
{D.A. Kirzhnits, On the internal structure of neutron stars, \em JETP,} {\bf 38}, (1960), 503 - 509.
\bibitem{TK_SVG}
{V.G. Soloviev, \em Theory of Atomic Nuclei: Quasiparticles and Phonons,} Institute of Physics, Bristol and Philadelphia, 1992.
\bibitem{TK_BK}
{N.B.	Brandt, V.A. Kulbachinskii, \em Quasiparticles in condensed matter physics,} ISBN 5922105647 (5-9221-0564-7), Hardcover, Fizmatlit, 2007.
\bibitem{TK_KL}
{M.I. Kaganov, I.M. Lifshits, \em Quasiparticles: ideas and principles of solid state quantum physics,} ISBN - 071471500X (Pbk), Moscow: Mir Publishers; London: Distributed by Central Books, 1980. 
\bibitem{TK_SSS} 
{E. Seres, J. Seres, C. Spielmann, X-ray absorption spectroscopy in the keV range with laser generated high harmonic radiation, \em Appl. Phys. Lett.,} {\bf 89} (2006),181919-181927.
\bibitem{TK_Gan}
{R.A.	Ganeev, Higher harmonics generation for intense laser radiation in plasma created by a prepulse acting on the surface of a solid target, \em UFN,} {\bf 179},  (2009), 65-90.
\bibitem{TK_Kit}
{Ch. Kittel, \em Introduction to Solid State Physics,} John Wiley and Sons, 2004. - ISBN 0-471-41526-X
\bibitem{TK_Tak1}
{N.Zh.	Takibayev, Class of Model Problems in Three-Body Quantum Mechanics That Admit Exact Solutions, \em Physics of Atomic Nuclei,} {\bf 71}, (2008), 460-468.
\bibitem{TK_Tak2}
{N.Zh.	Takibayev, Exact Analytical Solutions in Three-Body Problems and Model of Neutrino Generator, \em EPJ Web Conf.} {\bf 3}, (2010), 050281-050288; ArXiv: 1002.2257v1 [nucl-th]
\bibitem{TK_Tak3}
{N.Zh.	Takibayev, Neutron Resonances in Systems of Few Nuclei and Their
Possible Role in Radiation of Overdense Stars,  \em Few-Body Systems,} {\bf 50}, (2011), 311-314.
\end{thebibliography}
\end{document}